\documentclass[11pt]{article}
\usepackage{amsmath, amssymb, graphicx}
\usepackage{psfrag}
\usepackage[hyperindex=true,
          pdfstartview=FitH,
          bookmarksnumbered=true,
          bookmarksopen=true,
          citecolor=blue,
          linkcolor=blue,
          colorlinks=true, 
          pdfborder=001,   
          unicode]{hyperref}
\allowdisplaybreaks

\begin{document}

\title{C-metric like vacuum with non-negative cosmological constant in five dimensions}
\author{Liu Zhao { }and Bin Zhu\thanks{emails: {\it lzhao@nankai.edu.cn} and {\it binzhu7@gmail.com}}\\
Department of Physics, Nankai University, Tianjin 300071, P R China}
\date{}
\maketitle

\begin{abstract}
We present and analyze an exact 5-dimensional vacuum solution of Einstein equation with non-negative cosmological constant written in a C-metric like coordinate. The metric does not contain any black hole horizons in it, but has two acceleration horizons and a static patch in between. The coordinate system, horizon geometry and global structures are analyzed in detail, and in the case of vanishing cosmological constant, a simple exterior geometric interpretation is given. The metric possesses a spacelike Killing coordinate $\phi$ besides the timelike coordinate $t$, along which the spacetime can be dimensionally reduced via Kaluza-Klein mechanism and interpreted as Einstein gravity coupled to a 4 dimensional Liouville field (and a Maxwell
field as well if a boost operation is performed before the Kaluza-Klein reduction).
\end{abstract}

\section{Introduction}

Einstein gravity in higher dimensions admits unexpected rich structures due to the lack of uniqueness theorems. This fact has aided a lot for making a number of significant discoveries 
since ten years ago and henceforth, including the braneworld scenarios, novel types of black holes/rings/strings/branes etc. Usually, the richness of higher dimensional gravity is manifested through the appearance of exotic black objects which are absent in 4 dimensions (for recent reviews, see \cite{Emparan:2008p1729,Obers:2008p1707} and references therein) or through some novel compactification mechanisms \cite{Randall:1999p4414}. However, even on the level of empty spacetime without any black object living inside such richness already signify itself. In this paper we shall describe an exact vacuum solution of Einstein gravity with a cosmological constant in 5 dimensions. This solution does not contain any black hole in it, but the appearance of two acceleration horizons makes its global structure highly nontrivial. 

The metric to be studied will be presented in a C-metric like coordinate system. The C-metric is a 4 dimensional axial symmetric spacetime with two black holes accelerating apart. It is known since quite a long time ago \cite{L-Witten}, has found significant applications in constructing black ring solutions in 5 dimensions in the recent years \cite{Emparan:2001p607}, and admits very interesting physical features \cite{Griffiths:2006p3395, Hong:2003p3568, Dias:2003p2357}. In the light of the on-going intensive studies on higher dimensional gravities, it is a bit strange that no higher dimensional analogues of the C-metric has been found so far. It becomes clear only recently \cite{Podolsky-talk} that a higher dimensional analogue of the C-metric admitting black hole horizons does not exist at all. Therefore the best one can do in generalizing the C-metric to higher dimensions is to construct analogues of empty (i.e. all charges vanishing) C-metric spacetime in higher dimensions. The metric presented in this paper can be regarded as a preliminary result in this direction. Though the lack of black hole horizons makes the metric sound less interesting, the inclusion of a cosmological constant heals a bit to encode the metric with richer structures. 

Throughout this paper, the cosmological constant is considered to be non-negative. For negative cosmological constant, the metric is still an exact solution to the vacuum Einstein equation but the global structure will be different, and hence we shall not consider it here but leave it to a separate study. 

The paper is organized as follows. In Section 2, we shall present the metric to be studied with emphasis on the physically allowed ranges of the coordinate system. Section 3 is devoted to the studies of the horizon geometry. Then, in Section 4, we describe the global structure of the spacetime through the introduction of a number of coordinate transformations and by depicting the Penrose diagrams. In Section 5 we give a quick 4 dimensional interpretation of the metric via Kaluza-Klein (KK) reduction. The reduced metric corresponds to a 4 dimensional gravity coupled to a 4D Liouville field and possibly a Maxwell field. Section 6 studies the particular limit of $\Lambda=0$, i.e. vanishing cosmological constant. In this limit the spacetime admits an interesting exterior geometric description. The paper is ended in Section 7 with some concluding remarks. 

\section{The metric and coordinate ranges}

The metric we shall be studying in this article has the form
\begin{equation}
{ds}^2=\frac{1}{\alpha ^2(x+y)^2}\left[-G(y)H(z){dt}^2+G(y)\frac{{dz}^2}{H(z)}+\frac{{dy}^2}{G(y)}+\frac{{dx}^2}{F(x)}+F(x){d\phi}^2\right], \label{metric}
\end{equation}
where
\begin{align}
F(x)=1-x^2,\quad G(y)=-1-\frac{\Lambda }{6\alpha ^2}+y^2, \quad H(z)=1-\left(1+\frac{\Lambda }{6\alpha ^2}\right)z^2.
\end{align}
This is an exact solution to the 5-dimensional vacuum Einstein equation $R_{MN}-\frac{1}{2}g_{MN}R + \Lambda g_{MN}=0$ with cosmological constant
$\Lambda$. The coordinates
$(t, z, y, x, \phi)$ are intentionally taken such that the metric resembles very much to the famous C-metric in 4 dimensions. 

To interpret the metric as a physically acceptable spacetime we need to impose some constraint conditions over the ranges of coordinates. While doing so we kept the following principles in mind, i.e. 1) the spacetime must involve only one timelike direction and 2) there should be a region in the spacetime in which the metric becomes static. The existence of a static region is in fact not a physical requirement but a choice for convenience, which is enjoyed very much in the studies of usual de Sitter spacetime as well as 4-dimensional C-metric spacetime, so we impose it.

The metric (\ref{metric}) has two explicit Killing coordinates: $t$ and $\phi$. Either of these can play the role of a timelike coordinate (possibly after Wick rotation). For the moment we choose $t$ as timelike  and leave $\phi$ as an angular coordinate. 

Looking back on the metric (\ref{metric}), one can see that the overall conformal factor \(\frac{1}{\alpha ^2(x+y)^2}\) implies that the points satisfying
\(x+y=0\) form the conformal infinity. Therefore, the spacetime must lie on one side of \(x + y = 0\), i.e. we need to choose either \(x+y \ge 0\) or \(x+y \le 0\).
Without loss of generality, we make the first choice: \(x+y \ge 0\).

Each one of the functions \(F(x)\), \(G(y)\) and \(H(z)\) has two zeros
\begin{align}
x=\pm 1, \quad
y=\pm y_0, \quad 
z=\pm z_0, \label{y0z0}
\end{align}
where
\begin{align}
y_0 = \sqrt{1+\frac{\Lambda }{6\alpha ^2}},
\quad z_0 = \frac{1}{\sqrt{1+\frac{\Lambda }{6\alpha ^2}}}
=\frac{1}{y_0}. \label{y0z02}
\end{align}
For $\Lambda \ge 0$ we have
\begin{align*}
y_0 \ge 1, \quad z_0 \le 1,
\end{align*}
where the equalities hold only for $\Lambda=0$.

The requirement that only one timelike coordinate is present in the metric (1) implies that both \(F(x)\) and \(G(y)\) must always be non-negative. The requirement \(F(x)\ge 0\) 
implies that \(x\) must take values from \(-1\) to \(+1\), i.e. \(x \in [-1,1]\). This result, together with the requirements \(G(y)\ge 0\) and \(x+y\ge 0\), implies that \(y\) must take values in \( [y_0,\infty)\), and the other root \(-y_0\) of the function \(G(y)\) is beyond the physical region of the coordinates. Notice that for $\Lambda = 0$, there is a chance for the inequality \(x+y\ge 0\) to saturate, implying that the conformal infinity can be part of the spacetime; for $\Lambda >0$, \(y\) will always be bigger than 1 and hence it is impossible to saturate the inequality \(x+y\ge 0\), so the conformal infinity is well beyond the physical region of the spacetime. In other words, the spacial sections of the $\Lambda>0$ spacetime at a fixed time must have finite volume,
just like the usual maximally symmetric de Sitter spacetime.

Since \(x\in [-1,1]\), we can make a change of variable 
\(x= -\cos (\theta )\)
so that \(\theta\) takes values in \([0, \pi] \). Then the metric (\ref{metric}) becomes
\begin{align}
{ds}^2=\frac{1}{\alpha ^2(y-{\cos\theta })^2}\left[-G(y)H(z){dt}^2+G(y)\frac{{dz}^2}{H(z)}+\frac{{dy}^2}{G(y)}+{d\theta
}^2+\sin ^2\theta  {d\phi }^2\right]
\end{align}
Thus the geometry of the \((\theta ,\phi )\) directions is conformally equivalent to a 2-sphere. This helps in determining the physical range for the coordinate \(\phi\), which is taken as \(\phi \in [0, 2\pi)\).

If \(t\) is timelike then \(H(z)\) must be positive, i.e. \(z\in \left(-z_0, z_0\right)\). Conversely, if \(H(z)\) is negative then \(z\) will be timelike and we have either \(z\in (-\infty, z_0)\) or \(z\in(z_0,\infty)\). 
When \(H(z)>0\), the metric (\ref{metric}) is static. Otherwise it is non-static. The points \(z=\pm z_0\) are coordinate singularities of (\ref{metric}), which are causal boundaries between different spacetime regions (horizons). These horizons are not related to black holes -- they are only acceleration horizons, because all the curvature invariants are completely regular over the whole spacetime. For instance, the curvature scalar formed by contracting two Riemann tensors reads
\begin{align*}
R_{MNPQ}R^{MNPQ}
=\frac{10 \alpha^8 \Lambda^2}{9},
\end{align*}
which is everywhere regular.

To summarize, the coordinate ranges of our spacetime are given as follows:
\begin{align*}
t &\in (-\infty, \infty) ,\\
z &\in (-\infty, \infty) ,\\
y &\in [y_0 , \infty) ,\\
x &\in [-1 , 1] , \\
\phi &\in [0, 2 \pi) .
\end{align*}
\begin{figure}
\begin{center}
%\psfrag{A}{$A$}\psfrag{B}{$B$}\psfrag{C}{$C$}\psfrag{D}{$D$}
%\psfrag{E}{$E$}\psfrag{F}{$F$}\psfrag{G}{$G$}\psfrag{H}{$H$}
%\psfrag{M}{$M$}\psfrag{N}{$N$}\psfrag{P}{$P$}\psfrag{Q}{$Q$}
%\psfrag{x}{$x$}\psfrag{y}{$y$}\psfrag{z}{$z$}
\includegraphics[height=14cm]{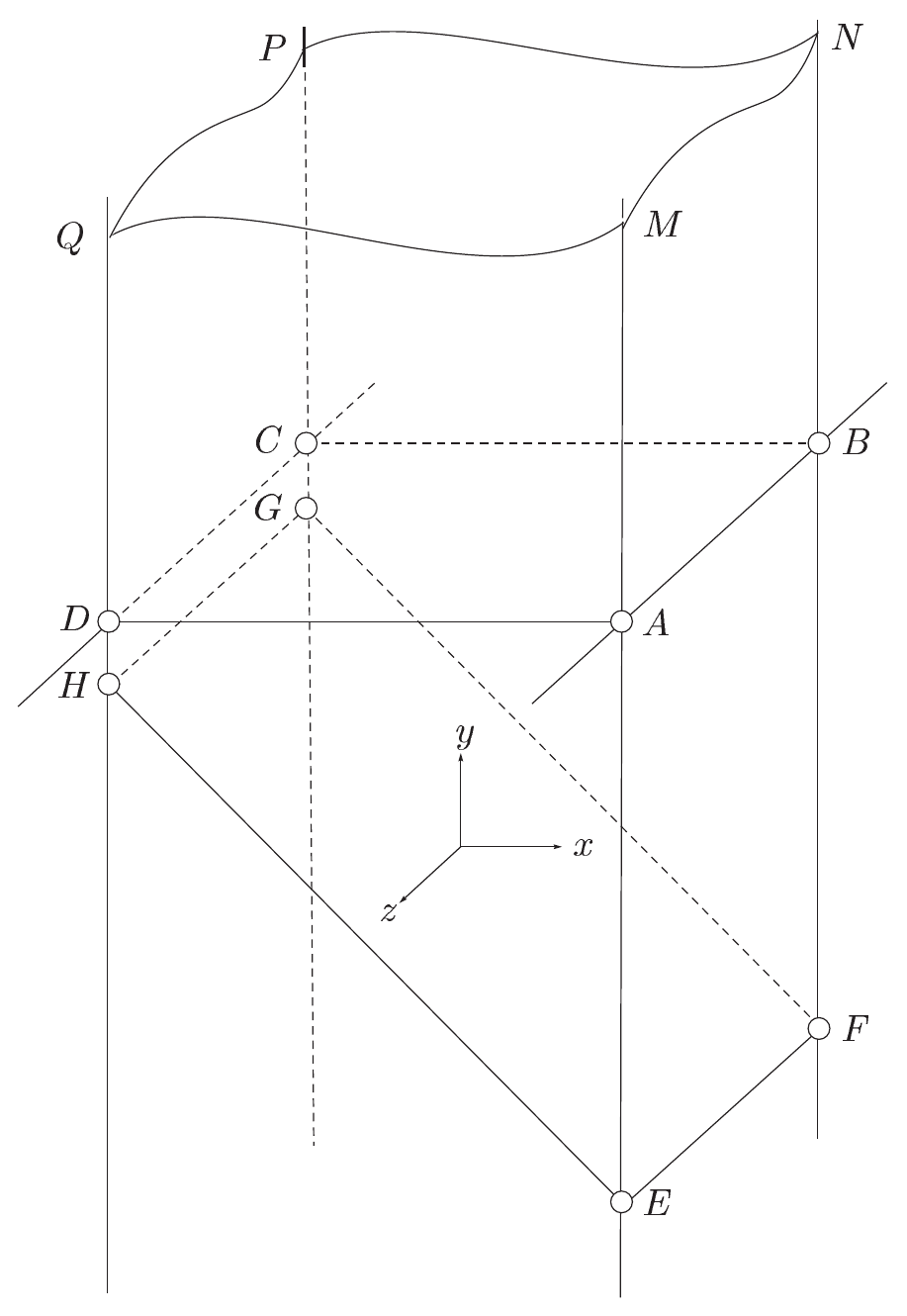}
\begin{minipage}{10.5cm}
\caption{The \((x,y,z)\) slice of the physical region of the spacetime: the static patch of the spacetime is the region bounded by the planes \((A,B,C,D)\), \((A,B,N,M)\), \((C,D,Q,P)\), \((A,D,Q,M)\) and \((B,C,P,N)\).}
\end{minipage}
\end{center}
\end{figure}

Figure 1 shows the \((x,y,z)\) slice of the physical region of the spacetime. In this figure, the \(x\) axis stretches from left to right, the \(y\) axis extends bottom up, and the \(z\) axis is transverse to the paper and extends from behind to the front of the paper sheet. The coordinate \(\phi\) is completely omitted and hence every point in the figure should be understood as a circle along \(\phi\). The \((x,y,x)\) coordinates of the twelve marked points are
\begin{align*}
&A: (1,y_0,z_0), \quad B:(1,y_0,-z_0), \quad  C: (-1,y_0,-z_0), \quad D: (-1,y_0,z_0)\\
&E: (1,-1, z_0), \quad F:(1, -1,-z_0), \quad  G: (-1, 1, -z_0), \quad H: (-1, 1, z_0)\\
&M: (1,\infty, z_0), \quad N: (1, \infty, -z_0), \quad  P: (-1, \infty, -z_0), \quad Q: (-1, \infty, z_0)
\end{align*}
The plane \((E,F,G,H)\) corresponds to \(x+y=0\), i.e. the conformal infinity. The plane marked with \((A,B,C,D)\) obeys \(y=y_0\), and the physical region of the spacetime is above this plane. Clearly the figure corresponds to the case \(\Lambda > 0\). For $\Lambda=0$, the lines $CD$ and $GH$ will overlap each other. The planes \((E,F,N,M)\) and \((G,H,Q,P)\) respectively correspond to \(x=\pm 1\), the maximal and minimal allowed values for \(x\), so the physical region is in between these two planes. The planar region \((A,D,Q,M)\) (with \(z=z_0\)) is one of the acceleration horizons. The other horizon \(z=-z_0\) is the planar region \((B,C,P,N)\). 
The static patch of the spacetime is the region bounded by the planes \((A,B,C,D)\), \((A,B,N,M)\), \((C,D,Q,P)\), \((A,D,Q,M)\) and \((B,C,P,N)\). Crossing the horizons the spacetime has two dynamic patches, namely the region in front of the plane \((A,D, Q, M)\) and bounded by the planes \((A,B,C,D)\), \((A,B,N,M)\), \((C,D,Q,P)\), and the region behind the plane \((B,C,P,N)\) and bounded by the planes \((A,B,C,D)\), \((A,B,N,M)\), \((C,D,Q,P)\).

\section{Horizon geometry} \label{hor}

Let us now consider the geometry of the horizons at \(z=\pm z_0\). The line element on (each of) the horizon reads
\begin{align}
ds_H^2=\frac{1}{\alpha ^2(y-{\cos\theta })^2}\left[\frac{{dy}^2}{y^2-y_0{}^2}+{d\theta }^2+\sin ^2\theta  {d\phi }^2\right]. \label{horizon}
\end{align}

In order to fully understand the geometry of the horizon, we need to make some further coordinate transformations. First let us make the change \(y\to \rho =\frac{1}{\alpha  y}\), where $\rho
$ takes value in the range
\begin{align*}
\rho \in \left[0, \rho _0\right], \quad \rho _0=\frac{1}{\alpha  y_0}
\end{align*}
After this change of variable, the metric for the horizon reads
\begin{align*}
{ds}_H^2=\frac{1}{(1-\alpha  \rho {\cos\theta })^2}\left(\frac{{d\rho }^2}{1-\left(\frac{\rho }{\rho _0}\right)^2}+\rho^2\left({d\theta }^2+\sin ^2\theta  {d\phi }^2\right)\right).
\end{align*}
Making a further change of coordinate
\(\rho  \to  \sigma = \arcsin \frac{\rho }{\rho _0}\), i.e. \(\rho =\rho _0\sin  \sigma \), we get
\begin{align*}
{ds}_H^2&=\frac{1}{\alpha ^2\left(y_0 -{\sin\sigma } {\cos\theta}\right)^2}\left({d\sigma }^2+\sin ^2\sigma  \left({d\theta }^2+\sin ^2\theta  {d\phi }^2\right)\right),
\end{align*}
where
\begin{align*}
\sigma  \in  \left[0,\frac{\pi
}{2}\right].
\end{align*}
This metric is clearly conformal to a 3-sphere provided \(y_0 >1\), i.e. \(\Lambda > 0\), but the conformal factor \(\frac{1}{\alpha^2\left(y_0-{\sin\sigma }
{\cos\theta }\right)^2}\) makes the local geometry differ from a standard 3-sphere
\footnote{Notice that the coordinate range for $\sigma$ is only half of the usual $3$-sphere.
This is because of the appearance of $\sin\sigma$ in the conformal factor.}.

What remains to ask is the area of the horizon. To calculate this, we just need to calculate the square root of the determinant of the horizon metric and integrate the result over the permitted range of coordinates on the horizon. The square root of the determinant of the horizon metric (\ref{horizon}) reads
\begin{align*}
\sqrt{g_{H}}=\frac{\sin\theta }{\alpha ^3(y-{\cos\theta })^3\left(y^2-y_0^2\right)^{1/2}}.
\end{align*}
Therefore, integrating over the whole horizon, we get the area of (each of) the horizon 
\begin{align*}
A=\int _0^{2\pi }d\phi \int _0^{\pi }d\theta \int _{y_0}^{\infty }dy\sqrt{g_{H}}=\frac{\pi ^2}{\left(y_0^2-1\right)^{3/2} \alpha^3},
\end{align*}
which is a finite number for \(y_0>1\) and is divergent for \(y_0=1\). Remembering the fact that \(y_0^2=1+\frac{\Lambda }{6\alpha ^2}\), we get
\begin{align*}
A=\pi ^2\left(\frac{6}{\Lambda }\right)^{3/2}.
\end{align*}
This area is equal to that of a 3-sphere of radius $r=(3/2\Lambda)^{1/2}$. 
As \(\Lambda \rightarrow 0\), this effective radius goes to infinity and that is why the horizon area diverges for \(\Lambda=0\). 

Since there is no black hole living inside the spacetime, we can only interpret the horizons as 
acceleration horizons or Rindler horizons. This can be justified by introducing the Rindler coordinate
\begin{align*}
\zeta =\frac{\left(G(y)H(z)\right)^{1/2}}{\alpha(x+y)} \qquad (\zeta > 0 )
\end{align*}
in the static region \(z\in (-z_0, z_0)\) so that the near horizon line element in the $t-z$ plane is changed into
\begin{align*}
{dl}^2=-\zeta ^2{dt }^2+{d\zeta }^2.
\end{align*}
Notice that the coordinate singularity at the horizons $z=\pm z_0$ are now transformed into $\zeta=0$. To study the nature of the horizons we must introduce Kruskal like coordinates which is nonsingular on the horizons. This is achieved via the coordinate transformations
\begin{align*}
X^- = -\zeta \exp(-t),\quad X^+ =\zeta \exp(t),
\end{align*}
in terms of which the Rindler line element becomes
\begin{align*}
{dl}^2=- d X^- d X^+.
\end{align*}
Remember that the coordinates \(X^\pm\) does not cover the whole spacetime but only the static region.

Now consider the proper acceleration of a particle with a timelike trajectory \(X^M(\tau)\) along the Killing vector field \(\xi =\partial_t\). In the coordinates \(X^\pm\), the Killing vector \(\xi\) can be rewritten as
\begin{align*}
\xi = X^+\partial_{X^+} - X^- \partial_{X^-},
\end{align*}
and it is easy to see that the horizons at \(X^+=0\) and \(X^-=0\) are normal to this Killing vector, i.e. they are Killing horizons with respect to \(\xi\). The proper velocity reads
\begin{align*}
u^M = \frac{\xi^M}{\left(-\xi^2\right)^{1/2}}.
\end{align*}
So, the corresponding proper acceleration is
\begin{align*}
a^M = D_\tau u^M = u^N \nabla_N u^M = \left(X^+\right)^{-1}\partial_{X^+}
+\left(X^-\right)^{-1}\partial_{X^-}.
\end{align*}
Finally, the amplitude of the proper acceleration is
\begin{align*}
|a| =\left( g_{MN} a^M a^N\right)^{1/2} = \left(-\frac{1}{X^-
X^+}\right)^{1/2} = \frac{1}{\zeta}.
\end{align*}
We see that \(|a|\) diverges on the horizons justifying the statement that they are acceleration horizons.

\section{Global structure of the spacetime}

To fully characterize the spacetime, it is necessary to identify the global structure thereof. To achieve this, let us look at the light rays
in the \(t-z\) plane, which are characterized by the condition
\begin{align*}
-H(z){dt}^2+\frac{{dz}^2}{H(z)}=0,
\end{align*}
i.e.
\begin{align*}
{dt}=\pm \frac{dz}{|H(z)|}.
\end{align*}
Integrating the above relation and name the result on the right hand side \(z^*\), we get
\begin{align*}
z^*=\frac{z_0}{2}\log \left|\frac{z_0+z}{z_0-z}\right|,
\end{align*}
in terms of which the light rays are expressed as
\begin{align*}
t\pm z^*=0.
\end{align*}
Lightlike variables in the new (tortoise) coordinate are given by
\begin{align*}
u=t-z^*,\quad v=t+z^*.
\end{align*}
The metric on the \(t-z\) plane is now written as\footnote{we use $dl$ to denote the ``radial'' line element in the \(t-z\) plane in order to distinguish from the complete line element $ds$.}
\begin{align*}
{dl}^2=-\frac{G(y)H(z)}{\alpha ^2(x+y)^2}{du} {dv}.
\end{align*}
The horizons at \(z=\pm z_0\) are now located at \(z^*=\pm \infty \) in terms of the tortoise coordinate. Thus the static patch of the spacetime is
infinitely large and we cannot describe the region of the spacetime beyond the horizons using \(z^*\). The way out is simple, though. In order
to analyze the global structure of the spacetime, we first need to make some conformal map of the coordinates such that at the horizons the value of the new coordinates are finite and the metric in the new coordinates is nonsingular. Then we can analytically extend across the horizons to get an effective description of the regions inside the horizons.
To proceed, however, we need to observe that the nonsingular coordinates cannot be defined identically from one horizon to the other.  So we need
to define nonsingular coordinates separately near each horizon.

Let us start from the static patch and consider the nonsingular coordinates near the horizon \(z=-z_0\) (or  \(z^*=-\infty \)). By the words ``near the horizon at $z=-z_0$'' we mean that we are considering part of the spacetime in the static patch for which $-z_0< z \le 0$. We introduce
\begin{align*}
U=-\exp \left(-\frac{u}{z_0}\right), \quad V=\exp \left(\frac{v}{z_0}\right),
\end{align*}
such that the region outside the horizon obeys \(U<0\), \(V>0.\)  The radial line element in this particular part of the spacetime reads
\begin{align*}
{dl}^2=-\frac{G(y)\left(z_0-z\right)^2}{\alpha ^2(x+y)^2}{dU} {dV},
\end{align*}
in which we have made use of the definition of the function $H(z)$. Introducing new coordinates \(T, R\) with relations
\begin{align*}
U=T-R,\quad V=T+R,
\end{align*}
the line element is turned into
\begin{align*}
{dl}^2=\frac{G(y)\left(z_0-z\right)^2}{\alpha ^2(x+y)^2}\left(-{dT}^2+ {dR}^2\right).
\end{align*}
It is easy to see that 
\begin{align*}
T^2-R^2=U V = -\frac{z_0+z}{z_0-z}.
\end{align*}
So the horizon \(z=-z_0\) corresponds to \(T=\pm R\), two straight lines in the \(T-R\) coordinates.  The last relation also implicitly defines
\(z\),  from which one sees that there is no singularity encountered while passing through the horizon. It should be noticed however, that after
crossing the horizon once, the sign of either \(U\) or \(V\) should change and the radial infinities at \(z=-\infty \) is located inside the horizon
at \(z=-z_0\), which obeys the relation
\begin{align*}
T^2-R^2=1.
\end{align*}
From regions where \(U>0, V>0\) or \(U<0, V<0\) we can cross the horizon once again to reach another region for which \(U>0\) and \(V<0\). This last
region is spacelike relative to where we started.

If we started from near the other horizon \(z=z_0\) (i.e. from inside the region $0\le z < z_0$)  instead, we could have defined the new coordinates \(\tilde{U}, \tilde{V}\) as
\begin{align*}
\tilde{U}=\exp \left(\frac{u}{z_0}\right), \quad \tilde{V}=-\exp \left(-\frac{v}{z_0}\right).
\end{align*}
Then, introducing further $\tilde{T}$ and $\tilde{R}$ via
\begin{align*}
\tilde{U}=\tilde{T}-\tilde{R},\quad \tilde{V}=\tilde{T}+\tilde{R},
\end{align*}
the line element could be changed into
\begin{align*}
{dl}^2=\frac{G(y)\left(z_0+z\right)^2}{\alpha ^2(x+y)^2}\left(-d\tilde{T}^2+ d\tilde{R}^2\right),
\end{align*}
where \(z\) is implicitly defined in terms of the following relation:
\begin{align*}
\tilde{T}^2-\tilde{R}^2= \tilde{U} \tilde{V} =-\frac{z_0-z}{z_0+z}.
\end{align*}
Clearly the horizon at \(z=z_0\) corresponds to \(\tilde{T}=\pm \tilde{R}\), and the spatial infinity \(z=\infty \) corresponds to
\begin{align*}
\tilde{T}^2-\tilde{R}^2=1.
\end{align*}
Once again, crossing the horizon does not encounter any singularity. Crossing once from near \(z=z_0\) means changing the sign of one of \(\tilde{U}\) and
\(\tilde{V}\), while crossing twice would result in changing the signs of both \(\tilde{U}\) and \(\tilde{V}\).

\begin{figure}[ht]
\begin{center}
%\psfrag{z=0}{$z=0$}\psfrag{z-}{$z=-\infty$}\psfrag{z+}{$z=\infty$}
%\psfrag{U}{$V$}\psfrag{V1}{$U$}
%\psfrag{U'}{$\tilde{V}$}\psfrag{V'}{$\tilde{U}$}
\includegraphics[width=14cm]{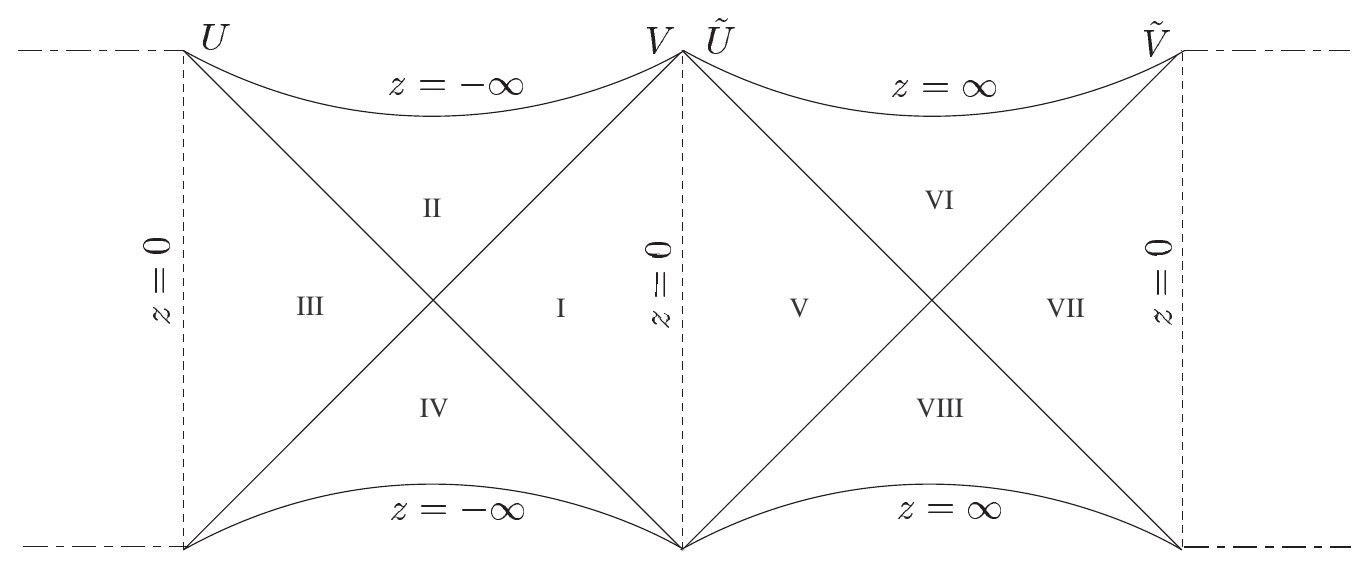}
\begin{minipage}{10.5cm}
\caption{
Penrose diagram of the spacetime: the \(T\) (\(\tilde{T}\)) axises point upwards, and the \(R\) (\(\tilde{R}\)) axises point towards right. Horizontal slashed lines represent repeated occurrences of the eight zones depicted in the middle.}
\end{minipage}
\end{center}
\end{figure}

The whole process for constructing the global structure of the spacetime is summarized in the Penrose diagram depicted in Figure 2. The diagram is in fact an infinite, periodic strip with 8 different zones in each period. The nonsingular, Kruskal-like coordinates systems are defined in each zone as follows:
\begin{itemize}
\item zone I:
\begin{align*}
U=-\exp \left(-\frac{u}{z_0}\right), \quad V=\exp \left(\frac{v}{z_0}\right);
\end{align*}
\item zone II:
\begin{align*}
U=\exp \left(-\frac{u}{z_0}\right), \quad V=\exp \left(\frac{v}{z_0}\right);
\end{align*}
\item zone III:
\begin{align*}
U=\exp \left(-\frac{u}{z_0}\right), \quad V=-\exp \left(\frac{v}{z_0}\right);
\end{align*}
\item zone IV:
\begin{align*}
U=-\exp \left(-\frac{u}{z_0}\right), \quad V=-\exp \left(\frac{v}{z_0}\right);
\end{align*}
\item zone V:
\begin{align*}
\tilde{U}=\exp \left(\frac{u}{z_0}\right), \quad \tilde{V}=-\exp \left(-\frac{v}{z_0}\right);
\end{align*}
\item zone VI:
\begin{align*}
\tilde{U}=\exp \left(\frac{u}{z_0}\right), \quad \tilde{V}=\exp \left(-\frac{v}{z_0}\right);
\end{align*}
\item zone VII:
\begin{align*}
\tilde{U}=-\exp \left(\frac{u}{z_0}\right), \quad \tilde{V}=\exp \left(-\frac{v}{z_0}\right);
\end{align*}
\item zone VIII:
\begin{align*}
\tilde{U}=-\exp \left(\frac{u}{z_0}\right), \quad \tilde{V}=-\exp \left(-\frac{v}{z_0}\right).
\end{align*}
\end{itemize}

\begin{figure}[ht]
\begin{center}
\includegraphics[width=8cm]{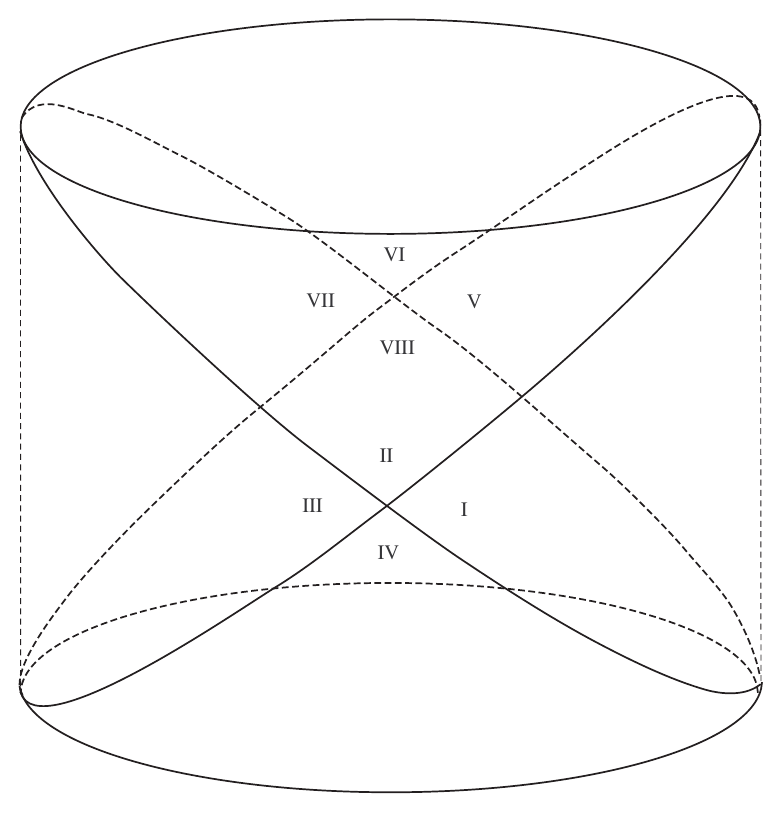}
\end{center}
\caption{Penrose diagram drawn on a cylindrical surface}
\end{figure}

To the left of zone III, there exists another zone VII${}'$ which look exactly like zone VII, and
to the right of zone VII, there exists another zone III${}'$ which look exactly like zone III. Hence the similar causal structure repeat horizontally and form an infinite periodic strip of Penrose diagram.

It may be advantageous to depict the Penrose diagram on a cylindrical surface by identifying the left boundary of zone III with the right boundary of zone VII (i.e. identifying zone III${}'$ with zone III, zone VII${}'$ with zone VII etc.). The result is shown in Figure 3. This way of depicting the Penrose diagram will make the periodicity of the Penrose diagram more manifest. From Figure 3, it can be seen that there two diamond-shaped radially bounded regions which are relatively spacelike to each other and both admit a static coordinate description. Each of these diamond shaped region (one consists of zone I and V, the other consists of zone III and VII) has a similar causal structure to that of the usual de Sitter spacetime. However, putting together the global structure of our spacetime is much richer than the usual de Sitter spacetime. 

\section{4D interpretation via Kaluza-Klein reduction}

The 5-dimensional metric (\ref{metric}) admits a 4-dimensional
interpretation via KK reduction.
Looking back at (\ref{metric}), we see that it can only be KK reduced along the Killing coordinate $\phi$ to avoid the appearance of massive KK modes. Since there is no off-diagonal element in the metric, the KK reduced theory will not contain any gauge field but with only a scalar Liouville field coupling to 4D Einstein gravity. The KK reduction formula reads\footnote{In this section we use the notation \(ds_5^2\) to denote the original line element in 5 dimensions in order to distinguish with the reduced 4 dimensional line element \(ds_4^2\).}
\begin{align}
ds_{5}^{2}=e^{\varphi/\sqrt{3}}ds_{4}^{2}+e^{-2\varphi/\sqrt{3}}d\phi^{2}, 
\label{red}
\end{align}
where $\varphi$ is a 4D scalar field. The 
4D metric and the scalar field $\varphi$ take the values
\begin{align}
ds_4^2 &= \frac{F(x)^{1/2}}{\alpha^3(x+y)^3}\left[-G(y)H(z){dt}^2+G(y)\frac{{dz}^2}{H(z)}+\frac{{dy}^2}{G(y)}+\frac{{dx}^2}{F(x)}\right],\label{zeroboost1}\\
e^{-2\varphi/\sqrt{3}} &= \frac{F(x)}{\alpha ^2(x+y)^2}.
\label{zeroboost2}
\end{align}

The KK reduction can be better understood on the level of actions. The action for the 5D vacuum Einstein equation $R_{MN}-\frac{1}{2}g_{MN}R + \Lambda g_{MN}=0$ is
\begin{align}
S_5 = \int d^5 x\sqrt{-g_{(5)}} \left( R_{(5)} - \Lambda \right)
\label{act}
\end{align}
up to possible boundary counter terms which we omit. After the KK reduction (\ref{red}), the action becomes
\begin{align*}
S_4 = \int d^4 x \sqrt{-g_{(4)}} \left( R_{(4)} - \frac{1}
{2}\left(  \partial \varphi \right)^{2}
- \Lambda e^{\varphi/\sqrt{3}}\right).
\end{align*}
Thus we see that the field $\varphi$ is a Liouville field in 4 dimensions and the 5D cosmological constant $\Lambda$ becomes the coupling constant of the 4D Liouville field.

It is tempting to make a boost in the \(t,\phi\) plane of the  metric (\ref{metric}) before doing the KK reduction. This will produce a Maxwell field in the reduced 4 dimensional theory. To do so, we first make the boost
\begin{align*}
t &\rightarrow T= t\cosh\beta - \phi \sinh\beta,\\
\phi &\rightarrow \Phi= - t \sinh \beta + \phi \cosh \beta.
\end{align*}
Inserting the above into the metric (\ref{metric}), we have
\begin{align*}
d\tilde{s}_5^2&=\frac{1}{\alpha ^2(x+y)^2}\left[
- \frac{G(y)H(z)-k^2 F(x)}{1-k^2}{dT}^2 +G(y)\frac{{dz}^2}{H(z)}+\frac{{dy}^2}{G(y)}+\frac{{dx}^2}{F(x)}\right.\\
& + \left.\frac{F(x)-k^2 G(y)H(z)}{1-k^2}{d\Phi}^2
+ \frac{2k( F(x)-G(y)H(z))}{1-k^2} dT d\Phi
\right],
\end{align*}
where we have changed the rapidity \(\beta\) into the boost velocity \(k\) via
\begin{align*}
k=\tanh\beta.
\end{align*}
Now making a KK reduction along the \(\Phi\) axis using the formula
\begin{align*}
d\tilde{s}_{5}^{2}=e^{\varphi/\sqrt{3}}d\tilde{s}_{4}^{2}
+e^{-2\varphi/\sqrt{3}}\left(d\Phi+\mathcal{A}\right)^{2}, 
\end{align*}
we get the reduced 4 dimensional metric
\begin{align}
d\tilde{s}_4^2&=\frac{1}{\alpha ^3(x+y)^3}
\left(\frac{F(x)-k^2 G(y)H(z)}{1-k^2}\right)^{1/2} \nonumber\\
&\times \left[- \frac{G(y)H(z)-k^2 F(x)}{1-k^2}{dT}^2 +G(y)\frac{{dz}^2}{H(z)}+\frac{{dy}^2}{G(y)}+\frac{{dx}^2}{F(x)}\right], \label{sol1}
\end{align}
together with the 4D Maxwell potential
\begin{align}
\mathcal{A}=\frac{k[  F(x)-G(y)H(z)]}{ F(x)-k^2 G(y)H(z)} dT
\end{align}
and the 4D Liouville field
\begin{align}
e^{-2\varphi/\sqrt{3}}= \frac{1}{\alpha ^2(x+y)^2}
\frac{F(x)-k^2 G(y)H(z)}{1-k^2}. \label{sol3}
\end{align}
Defining
\begin{align*}
F = F_{\mu\nu} dx^\mu dx^\nu \equiv d\mathcal{A},
\end{align*}
the reduced form of the action (\ref{act}) reads
\begin{align*}
\tilde{S}_4 &= \int d^4 x\sqrt{-g_{(4)}} \left( R_{(4)} - \frac{1}
{2}\left(  \partial \varphi \right)^{2}
- \Lambda e^{\varphi/\sqrt{3}} - \frac{1}{4} e^{\varphi/\sqrt{3}} F_{\mu\nu}F^{\mu\nu}\right),
\end{align*}
which is clearly an Einstein-Maxwell-Liouville theory. So we have obtained an exact solution to the Einstein-Maxwell-Liouville theory through the above process. It is easy to see that at $k=0$ the Maxwell field vanishes and the Einstein-Liouville solution (\ref{zeroboost1})-(\ref{zeroboost2}) is recovered.

\section{$\Lambda= 0$ and exterior geometry} \label{L0}

In this section we shall consider the limiting case $\Lambda= 0$ for the spacetime (\ref{metric}), i.e.
\begin{align}
{ds}^2=\frac{1}{\alpha ^2(x+y)^2}\left[-(y^2-1)(1-z^2){dt}^2+\frac{y^2-1}{1-z^2}{dz}^2
+\frac{{dy}^2}{y^2-1}+\frac{{dx}^2}{1-x^2}+(1-x^2){d\phi}^2\right]. \label{m2}
\end{align}
What is special about the $\Lambda= 0$ case? It is clear from Section \ref{hor} that the area of horizons now become infinite, signifying their noncompactness. Thus the case $\Lambda=0$ 
corresponds to a Ricci flat spacetime with two acceleration horizons, which may be identified as a Minkowski spacetime written in a particular accelerating coordinate system.  

To justify the last statement and further understand the meaning of the particular coordinate system chosen in writing (\ref{m2}), we will borrow some exterior geometric technology which can be found in \cite{Frolov:2006p310}. 

Instead of the metric (\ref{m2}), we will actually be investigating its Wick rotated version
\begin{align}
{ds}^2&=\frac{1}{\alpha ^2(x+y)^2}\left\{\frac{{dy}^2}{y^2-1}+(y^2-1)\left[\frac{{dz}^2}{1-z^2}+(1-z^2){d\psi}^2\right]\right.\nonumber\\
&\qquad\qquad\qquad +\left.\frac{{dx}^2}{1-x^2}+(1-x^2){d\phi}^2\right\}, \label{m2prime}
\end{align}
where $\psi$ has replaced the role of complexified time coordinate $it$.

Now consider the following rotational surface embedded in a 5-dimensional Euclidean space with coordinates \(X_i\) (\(i=1,2,...,5\)). The embedding equation reads
\begin{align}
X_1^2+X_2^2+\left(\sqrt{X_3^2+X_4^2+X_5^2}-a\right)^2=b^2. \label{embed}
\end{align}
For fixed constants $a>b$, this equation describes a compact 4-dimensional surface of topology 
$S^2\times S^2$. In fact, the surface described by the above equation can be thought of as the result of pulling the center of a 2-sphere of radius $b$ everywhere around another 2-sphere of radius $a$. 

We can parametrize the above surface in 5-dimensional Euclidean space as follows:
\begin{align}
X_1 &= \frac{\alpha}{B} \sin\theta\cos\phi,\label{p1}\\
X_2 &= \frac{\alpha}{B} \sin\theta\sin\phi,\\
X_3 &= \frac{\alpha}{B} \sinh\eta\sin\chi\cos\psi, \\
X_4 &= \frac{\alpha}{B} \sinh\eta\sin\chi\sin\psi,\label{p4}\\
X_5 &= \frac{\alpha}{B} \sinh\eta\cos\chi,\label{p5}
\end{align}
where
\begin{align*}
B &\equiv \cosh\eta - \cos\theta,\\
\alpha &\equiv \sqrt{a^2-b^2},
\end{align*}
provided $\eta$ takes the special value
\begin{align*}
\eta=\eta_0, \quad \cosh\eta_0 = \frac{a}{b}.
\end{align*}
For variable values of $\eta$, eqs. (\ref{p1})-(\ref{p5}) is just another parametrization of the 5-dimensional Euclidean space.

Making some further coordinate transform
\begin{align*}
x&=-\cos\theta,\\
y&=\cosh\eta,\\
z&=\cos\chi,
\end{align*}
it can be checked that the 5-dimensional Euclidean metric
\begin{align*}
ds^2= \sum_{i=1}^5 dX_i^2
\end{align*}
is equivalent to the metric (\ref{m2prime}). In other words, the space described by the metric (\ref{m2prime}) is just the 5-dimensional Euclidean space written in a subtle coordinate system.
In such a coordinate system, all constant $y$ hyper surfaces have the same $S^2\times S^2$ topology, however only the $y=a/b$ case corresponds to an ideal $S^2\times S^2$ local geometry.

Now looking back on eqs. (\ref{p1})-(\ref{p5}), we can see that the Wick rotation from $\psi$ to $it$ corresponds to Wick rotation of $X_4$, or, from the point of view of embedding surfaces,  corresponds to changing the equation (\ref{embed}) into 
\begin{align}
X_1^2+X_2^2+\left(\sqrt{X_3^2-X_4^2+X_5^2}-a\right)^2=b^2. \label{embedprime}
\end{align}
Therefore, following the same argument as above, the constant $y$ hyper surfaces in the metric
(\ref{m2}) are all topologically equivalent to $S^2\times H^2$, where $H^2$ is a 2-dimensional hyperbolic surface with equation $X^2+Y^2-Z^2=a^2$.

Before ending this section, let us remark that the global structure analysis made in Section 4 still holds for the $\Lambda=0$ case, the only difference lies in that we need to replace \(z_0\) everywhere by \(+1\). In addition, the KK reduction made in Section 5 also holds for \(\Lambda=0\), however for vanishing \(\Lambda\) the field \(\varphi\) is no longer a Liouville field but only dilaton field. Therefore (\ref{sol1})-(\ref{sol3}) correspond to a solution to the Einstein-Maxwell-dilaton theory for vanishing \(\Lambda\).

\section{Discussions}

The metric (\ref{metric}) studied in this paper turns out to be much more interesting than it 
first appears to be. The existence of the two acceleration horizons makes the global structure 
of the spacetime quite nontrivial. To make comparisons, let us remind that the usual de Sitter spacetime has only one acceleration horizon which is spherically symmetric with constant scalar curvature. Now the metric (\ref{metric}) processes {\em two} disjoint acceleration horizons and the scalar curvatures of both horizons are not constants. The de Sitter C-metric \cite{Dias:2003p2357} with
vanishing mass and charges in 4 dimensions do have two acceleration horizons but the two horizons in that case are located in two causally disconnected regions in the spacetime, so that each observer in that spacetime can perceive the existence of only one of them. However the two acceleration horizons in (\ref{metric}) can be perceived from a single causal patch of the spacetime.

Moreover, via KK reductions, the metric can be reduced into an exact solution for the Einstein-Maxwell-Liouville/dilaton theory, which is of course nontrivial in 4 dimensions.

The rich global structure certainly deserves further explorations.  It provides one further example for the richness of higher dimensional gravities besides existing ones on the one hand, and it also indicates the unexpectedly many varieties of Einstein manifolds on the other. The metric (\ref{metric}) represents an empty, non-spherical symmetric Einstein manifold. Traditionally more familiar examples of such manifolds are de Sitter spacetime (for \(\Lambda>0\)) or Minkowski spacetime (for \(\Lambda=0\)). It is interesting to ask how many different varieties of such manifolds exist in higher dimensions. We do not have an answer for this question at the moment, but we do have some other explicit examples of this kind which is similar to but essentially different from the metric (\ref{metric}). One such example is
\begin{align}
{ds}^2 = \frac {1} {\alpha^2(x + y)^2}
\left[ - G (y)dt^2 + \frac { {dy}^2} {G (y)}+ \frac {{dx}^2} {F (x)} + 
F (x)\left (\frac { {dz}^2} {H (z)} + 
H (z)d\phi^2 \right) \right], \label{newmetric}
\end{align}
with
\begin{align*}
F (x) = 1 - x^2, \quad G (y) = -\frac {\Lambda} {6\alpha^2} - 1 + y^2, \quad H (z) = 1 - z^2.
\end{align*}
This is also an exact solution to the 5-dimensional vacuum Einstein equation $R_{MN}-\frac{1}{2}g_{MN}R + \Lambda g_{MN}=0$ with cosmological constant $\Lambda$. However, as we can see from the above metric, it is \(y\) that is playing the role of radial coordinate in stead of \(z\). There are two acceleration horizons located at \(y=\pm y_0\). Moreover, the geometry of the \((x,z,\phi)\) slice is a conformal 3-sphere which can be seen by the transformation \(x=-\cos\theta_1, z=\cos\theta_2\). 
Therefore, the metric (\ref{newmetric}) is more symmetric in some sense. It would be interesting to explore the global structure of the metric (\ref{newmetric}) in a similar fashion as did in this paper. Also, the metric (\ref{newmetric}) may be used to construct exact solutions to the 4 dimensional Einstein-Maxwell-Liouville/dilaton theory via KK reduction.

What is intentionally left untouched in this paper is the \(\Lambda<0\) case of the metric. This case is much more complicated than the \(\Lambda\ge 0\) case. Indeed, from (\ref{y0z02}) it is clear that \(\Lambda\) cannot be arbitrarily negative in order that the acceleration horizons exist. The minimum allowed value of \(\Lambda\) is \(-6\alpha^2\). For any allowed negative value of \(\Lambda\), the rectangles \((A,B,C,D)\) and \((E,F,G,H)\) in Figure 1 will intersect each other, and the physical region of the coordinates used is changed drastically. We leave the detailed description for the \(\Lambda<0 \) case to somewhere else.

\section*{Acknowledgment} 

This work is supported by the National  Natural Science Foundation of China (NSFC) through grant No.10875059.

%\bibliographystyle{utcaps}
%\bibliography{papers,non-arxiv-refs}

\providecommand{\href}[2]{#2}\begingroup\raggedright\endgroup

\end{document}